# On the effect of linear feedback and parametric pumping on a resonator's frequency stability

**Zohreh Mohammadi[1], Toni L. Heugel[2], James M.L. Miller[3], Dongsuk D. Shin[3], Hyun-Keun Kwon[3], Thomas W. Kenny[3], Ramasubramanian Chitra[2], Oded Zilberberg[2], Luis Guillermo Villanueva[1]**

[1] Advanced NEMS Laboratory, Ecole Polytechnique Fédérale de Lausanne (EPFL), CH-1015, Lausanne, Switzerland
[2] Institute for Theoretical Physics, ETH Zurich, 8093 Zurich, Switzerland
[3] Departments of Mechanical and Electrical Engineering, Stanford University, Stanford, California 94305, USA

E-mail: Guillermo.Villanueva@epfl.ch

## Abstract

Resonant sensors based on Micro- and Nano-Electro Mechanical Systems (M/NEMS) are ubiquitous in many sensing applications due to their outstanding performance capabilities, which are directly proportional to the quality factor ($Q$) of the devices. We address here a recurrent question in the field: do dynamical techniques that modify the effective $Q$ (namely parametric pumping and direct drive velocity feedback) affect the performance of said sensors? We develop analytical models of both cases, while remaining in the linear regime, and introduce noise in the system from two separate sources: thermomechanical and amplifier (read-out) noise. We observe that parametric pumping enhances the quality factor in the amplitude response, but worsens it in the phase response on the resonator. In the case of feedback, we find that $Q$ is enhanced in both cases. Then, we establish a solution for the noisy problem with direct drive and parametric pumping simultaneously. We also find that, in the case when thermomechanical noise dominates, no benefit can be obtained from neither artificial $Q$-enhancement technique. However, in the case when amplifier noise dominates, we surprisingly observe that a significant advantage can only be achieved using parametric pumping in the squeezing region.

## 1. Introduction

Resonant-based sensors are widely used in our society. Their working principle is based on detecting the shift of the resonance frequency caused by an external effect. Indeed, this detection scheme is preferred over its static counterpart due to the gain in responsivity and/or resolution resulting from the high quality factor of the resonator. It is for this latter point that mechanical resonators are particularly interesting for sensing applications, as typical quality factors are larger than in the case of, e.g., LC resonators. In addition, reducing the size of mechanical devices also improves their responsivity to many kinds of phenomena. The combination of small size and large quality factor has enabled extreme resolutions when measuring, for example, mass[1], gas[2] or liquid[3] concentration, force[4], biological entities[5], charge[6], temperature[7] and radiation[8-9].

Reducing the size of the resonators is challenging as it is directly correlated with reducing the quality factor[10]. Therefore, with the advent of Micro- and Nano-ElectroMechanical Systems (MEMS/NEMS), much work has been directed towards artificially improving the quality factor: On the one hand, using passive approaches, including acoustic reflectors[11], geometry optimization[12], surface treatments[13] and dissipation dilution due to intrinsic stress[14]. All of them genuinely increase the quality factor, thus reducing the amount of thermomechanical noise (in force) allowed to enter the system; On the other hand, we have the so-called active approaches, mainly parametric pumping (tuning the resonance frequency at twice the rate of said frequency)[15-16] and feedback proportional to the velocity[17].

The motivation to improve the quality factor stems from the well-known Robbins formula [18]: In the canonical example of a resonator which frequency is tracked by means of a phase locked loop (PLL) [1, 19], the minimum detectable frequency change is determined by the noise in the determination of the resonator's phase divided by the slope of the phase-*vs.*-frequency curve close to the resonance frequency. As this latter term is the inverse of the linewidth, the minimum relative frequency shift that can be detected is given by $\sim \frac{S_\varphi^{1/2}}{Q}$, with $Q$ being the quality factor and $S_\varphi^{1/2}$ the noise in the determination of the resonator's phase.

In this paper, we calculate both the slope of the phase-*vs.*-frequency curve and $S_\varphi^{1/2}$ for the two mentioned cases of active modification of $Q$, i.e., parametric pumping (thus retaking Cleland's seminal work [20]) and positive feedback. We perform this calculation considering systems limited not only by thermomechanical but also by amplifier noise. In order to simplify our analysis, we consider the case of a linear resonator whose amplitude is limited by a given *onset of nonlinearity* (or critical amplitude). This is a very common practice to simplify the analysis of noise in resonators and it fundamentally differs from other approaches to improve frequency resolution by using nonlinearity in the system [21-22]. To avoid loss of generality, no particular source of nonlinearity is considered, and the only hard condition is that the onset of nonlinearity depends on the original linewidth, not the one actively modified.

The paper starts with a brief derivation of the equations describing the dynamics of the systems under study, which are applied to the case of parametrically-pumped and feedback-driven systems. The slope of the phase-vs.-frequency curves and the noise in the phase are presented in each case as a function of the strength and phase of the parametric pump and direct feedback. We conclude by clarifying once and for all the effect of these active $Q$ enhancement techniques on the frequency shift resolution.

## 2. Modeling, equations, and solution methods

The system under study is a linear damped resonator that is driven via a harmonic force close to the natural frequency ($\omega_0$) of the resonator and a noise term (thermomechanical noise) that has a white power spectral density. On top of that noise source, we also consider the so-called *amplifier noise* which in principle does not enter the equation of motion as it only affects the determination of the displacement by the measurement technique.

Two additional terms are included in the equation of motion (e.o.m.): a parametric pumping term which modulates the elastic constant of the resonator at a given rate ($\omega_p$ close to twice the natural frequency), and a feedback term proportional to the velocity of the resonator with a certain phase shift. To solve this system, we start by converting the original e.o.m.

$$m \frac{d^2 \tilde{x}}{d\tilde{t}^2} + \Gamma \frac{d\tilde{x}}{d\tilde{t}} + m\omega_0^2 \tilde{x} + \widetilde{H} \sin(\widetilde{\omega}_P \tilde{t} + \phi)\,\tilde{x} \\ + \tilde{\gamma}\frac{d\tilde{x}(\tilde{t}+\tilde{\psi})}{d\tilde{t}} = \widetilde{G} \cos(\widetilde{\omega}_D \tilde{t}) + \widetilde{F}_{th}(\tilde{t}) \quad (1)$$

into a dimensionless form [23]. In Eq. (1), $m$ is the effective mass, $\Gamma$ is the linear damping rate, $\widetilde{G}$ is the harmonic driving force at a frequency $\widetilde{\omega}_D$, $\widetilde{H}$ is the parametric pumping term that modulates the elastic constant with a phase delay $\phi$ with respect to the drive, $\tilde{\gamma}$ is the feedback term accounting for the magnitude, and $\tilde{\psi}$ its time delay. The term $\widetilde{F}_{th}$ is the thermomechanical noise (white and Gaussian) with $\langle \widetilde{F}_{th}(\tilde{t})\widetilde{F}_{th}(\tilde{t}')\rangle = \sigma_F^2 \delta(\tilde{t}-\tilde{t}')$, $\tilde{t}$ is the time, and $\tilde{x}$ is the displacement of the resonator mode at the maximum position for a given mode. Eq. (1) develops into its dimensionless form (Eq. (2)) after applying that $Q^{-1} = \Gamma/m\omega_0$; $t = \omega_0 \tilde{t}$; $x = \tilde{x}/x_c$; $|\gamma| = \tilde{\gamma}/\Gamma = \tilde{\gamma}Q/m\omega_0$; $\psi = \tilde{\psi}\omega_0$; $|g| = Q^{3/2}\widetilde{G}/m\omega_0^2 x_c$; $|h| = Q\widetilde{H}/2m\omega_0^2$; $\Omega_D = (\widetilde{\omega}_D/\omega_0 - 1)Q$; $\Omega_P = (\widetilde{\omega}_P/\omega_0 - 2)Q$; $\xi_{th}(t) = Q^{3/2}\widetilde{F}_{th}(\tilde{t})/m\omega_0^2 x_c$; $\langle \xi_{th}(t)\xi_{th}(t')\rangle = \sigma_\xi^2 \delta(t-t')$; $\sigma_\xi = Q^{3/2}\sigma_F/m\omega_0^{3/2} x_c$

$$\ddot{x} + \frac{1}{Q}\left(\dot{x} + |\gamma|\dot{x}(t+\psi)\right) \\ + \left(1 + 2\frac{|h|}{Q}\sin\left(\left(2+\frac{\Omega_P}{Q}\right)t + \phi\right)\right)x \\ = \frac{|g|}{Q^{3/2}}\cos\left(\left(1+\frac{\Omega_D}{Q}\right)t\right) + \frac{\xi_{th}(t)}{Q^{3/2}} \quad (2)$$

where the dot stands for derivative with respect to the dimensionless time variable, $t$. Assuming now that the quality factor of the resonator is high ($Q \geq 100$), we insert $x(t) = Q^{-\frac{1}{2}}|A(T)|\cos(t+\varphi)$ with $T = t/Q$, and use secular perturbation theory[23] with noise[24-25] to obtain the following *amplitude equation*:

$$\frac{dA(T)}{dT} + \frac{1+\gamma}{2}A(T) - \frac{h}{2}A^*(T)e^{i\Omega_P T} + \frac{i}{2}g e^{i\Omega_D T} \\ + \Xi_{Th}(T)e^{i\Omega_D T} = 0, \quad (3)$$

where $g = |g|$, $h = |h|e^{i\phi}$, $\gamma = |\gamma|e^{i\psi}$ and $\Xi_{Th}(T) = \frac{e^{-i\Omega_D T}}{2\pi}\int_{QT-\pi}^{QT+\pi}\xi(t)e^{-it}dt$. The noise $\Xi_{Th}(T)$ can be split into real $\Xi_R(T)$ and imaginary part $\Xi_I(T)$ with $\langle \Xi_R(T)\Xi_R(T')\rangle = \langle \Xi_I(T)\Xi_I(T')\rangle \approx \sigma_\Xi^2 \delta(T-T')$ and $\sigma_\Xi = \frac{\sigma_\xi}{\sqrt{2Q}}$. Eq. (3) can be used to calculate the amplitude $A(T)$, which is a variable that describes the slow deviations from the purely periodical response of the system, i.e., the complex amplitude within the rotating frame. As Eq. (3) has noise, the solution can be split into first finding the coherent response to the drives and later finding the power spectral density of the noise around said coherent response.

*2.1 Coherent response*

The solution to an equation of the type of Eq. (3) is given in general by $A(T) = a'e^{i\Omega_D T} + b'e^{i\Omega_P T}$, but for the sake of simplicity we here restrict ourselves to the so-called *degenerate* case, which happens when the pump frequency exactly doubles the drive frequency, namely $A(T) = ae^{i\Omega_D T}$, where $a = |a| \cdot e^{i\varphi}$ is a time independent complex variable. Substituting this ansatz into the coherent part of Eq. (3), we obtain:

$$2\Omega_D a - i(1+\gamma)a + iha^* = -g. \quad (4)$$

Taking now into account that $h$, $\gamma$ and $a$ are complex variables, we can write:

$$\frac{a}{|g|} = \frac{-2\Omega_D - i + i|h|e^{i\phi} - i|\gamma|e^{-i\psi}}{1 - |h|^2 + |\gamma|^2 + 4\Omega_D^2 + 2|\gamma|\cos(\psi) + 4|\gamma|\Omega_D \sin(\psi)} \quad (5)$$

Eq. (5) is similar to previous derivations that can be found in the literature[23], combining both linear feedback and parametric pumping[26]. Here we now consider independently both of those cases in order to compare them. This has, to the extent of the authors' knowledge, never been done before (cf. Figure 1).

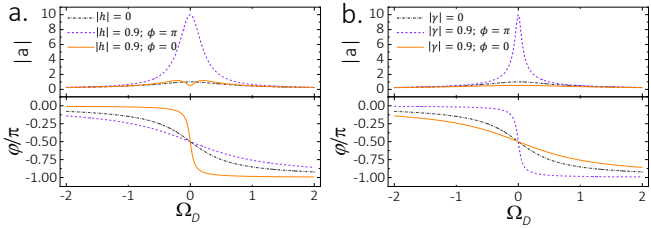

Figure 1 Comparison between a parametrically-pumped response and a response with linear feedback. | (a) Amplitude and phase response as a function of frequency for a directly driven system with no pump (dashed, black), and with large pump with two different relative delays between the direct drive signal and the pump. These two different delays correspond to the condition for maximum (dotted, purple) and minimum amplitude gain (solid, orange). The corresponding phase plots are shifted so that they cross at the natural frequency. (b) The same model as in (a), with linear feedback instead of parametric pumping. Both cases are very similar for the amplitude response. In contrast, the phase response presents a counter-intuitive behavior: a large gain parametric pump corresponds to a small phase slope and vice-versa.

We first look into the amplitude response as a function of frequency (top panel in Fig. 1). We observe that both the linear feedback and the parametrically pumped case have a given phase ($\phi = \psi = \pi$) for which the system presents a "gain" and a different phase phase ($\phi = \psi = 0$) for which the system yields an "attenuation", when the drive frequency is at the natural frequency of the device. When linear feedback is applied, it can be seen that the "gain" case ($\psi = \pi$) shows a phase response with an increased slope around the resonant frequency, and the opposite is true for the "attenuation" case.

However, when the system is parametrically pumped, the "gain" case in the amplitude response corresponds to a *reduction* of the phase slope close to the resonant frequency; and vice-versa. While the result shown in Figure 1.b for the system with feedback is rather intuitive, we find that the result in Figure 1.a has not been deeply analyzed before and exhibits a counter-intuitive behavior. We also want to highlight the fact that the way we define the parametric pumping and its phase allows us to find the "attenuation" and "gain" cases for $\phi = 0 \ \& \ \pi$ respectively, as opposed to the more commonly used values of $\pm\pi/4$.

### 2.2. Noisy solution

In the previous section, we have dismissed the thermomechanical noise driving term when proceeding from Eq. (3) to Eq. (4). As soon as that noise term is considered, the solution turns into $a + \delta a(T)$ with the second term being time dependent as corresponding to the amplitude response to $\Xi_{Th}(T)$. In order to solve for $\delta a(T)$, we need solve the following equation

$$-2i\dot{\delta a} + 2\Omega_D \delta a - i(1+\gamma)\delta a + ih\delta a^* = -2i\Xi_{Th}, \quad (6)$$

where $\Xi_{Th} = \Xi_R + i\Xi_I$, with $\Xi_R$ and $\Xi_I$ being the two (uncorrelated) components of the thermomechanical noise. As $\Xi_R$ and $\Xi_I$ are noisy variables (white, Gaussian), it is possible now to calculate the probability density function (PDF) of the amplitude $a + \delta a$, using a Fokker-Planck approach. We, then, represent the solutions in Figure 2 for $\Omega_D = 0$ and separately for the case of linear feedback (Figure 2.b) and parametric pumping (Figure 2.a).

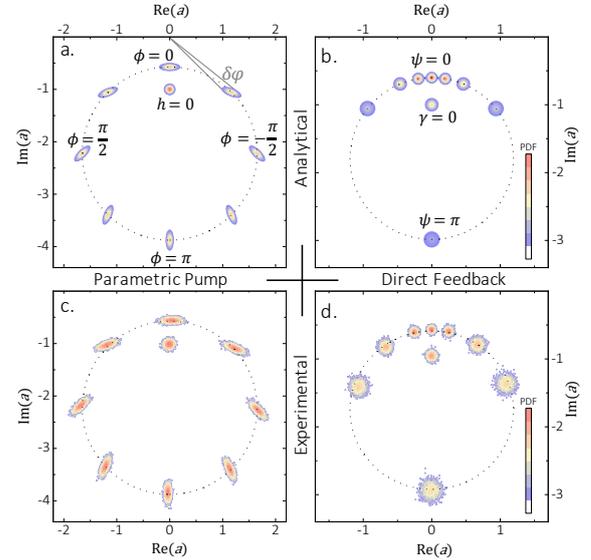

Figure 2 Noisy solution | (a) Analytical and (c) experimental PDF for $\Omega_D = 0$ and $|h| = 0.7$ for different values of $\phi$ in the case of parametric pumping. Squeezing can be observed in all cases, but in different directions in the rotatory space. (b) Analytical and (d) experimental PDF for $\Omega_D = 0$ and $|\gamma| = 0.7$ for different values of $\psi$ in the case of linear feedback. No squeezing occurs in this case.

The differences between both cases are striking. In the former case, Figure 2.b shows that by modifying the feedback *delay* (represented in our case by the phase $\psi$), it is possible to go from *active cooling* to *amplification*[17], but in all cases the noise is distributed isotopically around the mean value of the amplitude. In contrast, Figure 2.a also shows variation of the amplitude but, importantly, the noise distribution is not isotropic around the mean values and exhibits the well-known squeezing phenomenon instead. What is more interesting is the implications that this assymetry has on the determination of the modulus and phase of the complex amplitude $a + \delta a$. Figure 2.b has been presented in different venues before, but this is the first time that Figure 2.a is shown (the noisy solution of parametric pumping together with direct drive).

In Figure 2.c and Figure 2.d, we plot a collection of experimental data obtained in analogous conditions to those used in Figure 2.a and Figure 2.b, respectively. The experimental system consists of a microcantilever fabricated within a wafer-scale encapsulation process, utilizing internal electrodes and a low-noise capacitive readout to detect the thermomechanical vibrations with minimal added noise. This is an ideal platform for the study of the effect of noise, feedback and parametric pumping on the dynamical response of a resonant system[27-28].

Interestingly, the noise in the determination of the phase of the resonator, $\delta\varphi$, depends strongly on the phase and magnitude of the pump and feedback, respectively. To describe the representation of Figure 2 with formulas, we can calculate the power spectral density $S_{\delta a,R}$ and $S_{\delta a,I}$ of both the real and imaginary part of $\delta a$:

$$S_{\delta a,R} = I_{th} \frac{1 + |h|^2 + |\gamma|^2 + 4\Omega_D^2 + 2(|\gamma|\cos(\psi) + |h|\cos(\phi)) + 2|h||\gamma|\cos(\psi - \phi) + 4\Omega_D(|\gamma|\sin(\psi) + |h|\sin(\phi))}{(1 - |h|^2 + |\gamma|^2 + 4\Omega_D^2 + 2|\gamma|\cos(\psi) + 4|\gamma|\Omega_D\sin(\psi))^2}, \quad (7)$$

$$S_{\delta a,I} = I_{th} \frac{1 + |h|^2 + |\gamma|^2 + 4\Omega_D^2 + 2(|\gamma|\cos(\psi) - |h|\cos(\phi)) - 2|h||\gamma|\cos(\psi - \phi) + 4\Omega_D(|\gamma|\sin(\psi) - |h|\sin(\phi))}{(1 - |h|^2 + |\gamma|^2 + 4\Omega_D^2 + 2|\gamma|\cos(\psi) + 4|\gamma|\Omega_D\sin(\psi))^2} \quad (8)$$

where we have used that $\Xi_R$ and $\Xi_I$ are approximately uncorrelated and that their power spectral densities are: $I_{th} = \langle\Xi_R(\omega_S)\Xi_R(\omega_S')\rangle = \langle\Xi_I(\omega_S)\Xi_I(\omega_S')\rangle \approx \sigma_\Xi^2$. Importantly, the magnitude of $I_{th}$ is related to temperature, the quality factor, mass, and frequency of the resonator. However, neither parametric pumping nor linear feedback actually change this magnitude, i.e., they only change the effect of this noise in the amplitude [29].

From Eqs. (7) and (8) it is difficult to extract qualitative information but we can simplify these formulas considering our two separate cases (parametric pumping and linear feedback), and by using $\Omega_D \approx 0$. The latter approximation is made for the sake of simplicity and because sensors typically will satisfy this condition as the integration times tend to be long relative to the decay rate of the resonator. Crucially, however, the final conclusions of the paper remain essentially the same also when we do not take this approximation. Taking all assumptions described before and the limits $\gamma \to 0$ and $h \to 0$, we can obtain respectively Eqs. (9) and (10).

$$\begin{cases} S_{\delta a,R}(\Omega = 0) = I_{th} \frac{1 + |h|^2 + 2|h|\cos(\phi)}{(1 - |h|^2)^2} \\ S_{\delta a,I}(\Omega = 0) = I_{th} \frac{1 + |h|^2 - 2|h|\cos(\phi)}{(1 - |h|)^2} \end{cases}; \quad (9)$$

$$\begin{cases} S_{\delta a,R}(\Omega = 0) = \frac{I_{th}}{1 + |\gamma|^2 + 2|\gamma|\cos(\psi)} \\ S_{\delta a,I}(\Omega = 0) = \frac{I_{th}}{1 + |\gamma|^2 + 2|\gamma|\cos(\psi)} \end{cases}. \quad (10)$$

Eq. (10) clearly illustrates the isotropic distribution that can be seen in Figure 2.b&d, whereas Eq. (9) yields an anisotropic result that is dependent on the phase delay between drive and pump.

## 3. Phase and Frequency noise comparison

Our main objective in this paper is to compare the performance of a resonator as a resonant-based sensor when either feedback or parametric pumping is used to modify the dynamical response of the resonator. To do this, we need to first determine the noise in the resonator's phase, which is typically defined by: $S_\varphi = \frac{S_{\delta a}}{|a|^2}$. Then, as it is already mentioned in the introduction, one can make an estimation of how much noise is in the determination of the frequency based on: $S_\Omega = S_\varphi / \left(\frac{\partial\varphi}{\partial\Omega}\right)^2_{\Omega=0}$. This method can be easily applied to the case when $h = 0$. When a parametric pumping is present, however, each of the terms determining the noise in frequency depends on the relative phase between drive and pump, $\phi$. In fact, the slope of the phase $\frac{\partial\varphi}{\partial\Omega}$, the amplitude $|a|$, and the projection of the noise $\delta a$ in quadrature with $a$ all depend on $\phi$. The latter point is particularly interesting as it is where our conclusion differs from that of Cleland's [20]:

$$S_\varphi = \left(S_{\delta a,R}\left(\frac{1 - |h|\cos(\phi)}{\sqrt{1 + |h|^2 - 2|h|\cos(\phi)}}\right)^2 + S_{\delta a,I}\left(\frac{|h|\sin(\phi)}{\sqrt{1 + |h|^2 - 2|h|\cos(\phi)}}\right)^2\right) / |a(\phi)|^2 \quad (11)$$

as opposed to $S_\varphi = (S_{\delta a,R} + S_{\delta a,I})/2|a(\phi)|^2$.

In the following figures, Figure 2, Figure 2 and Figure 2, we show the effect of parametric pump and direct feedback on: (i) the noise in the resonator's phase, (ii) the resonator's phase response slope, and (iii) the resultant estimated frequency noise; always compared to the case without any feedback or parametric pumping.

In the first case, we see in Figure 2 the effect on the system where we keep the driving amplitude $|g|$ constant and the noise is fully thermomechanical. In the case of direct feedback, we can see how the noise in the resonator's phase ($S_\varphi$) is constant in all cases, and how the phase slope $\left(\frac{\partial \varphi}{\partial \Omega}\right)$ increases in the proximity of $\psi = \pi$. Thus, the overall frequency noise ($S_\Omega$) reduces, as expected. The situation is different in the case of parametric pumping, where the noise in the phase changes considerably depending on the delay between the pump and the drive, observing a minimum around $\phi = \pi$. Interestingly, as shown in Figure 1, the slope of the phase is reduced at that value and is maximized for $\phi = 0$. The final result for the frequency noise shows that the optimum operational point would be $\phi = \pi$, because the reduction in the noise in the phase is larger than the reduction in the phase slope.

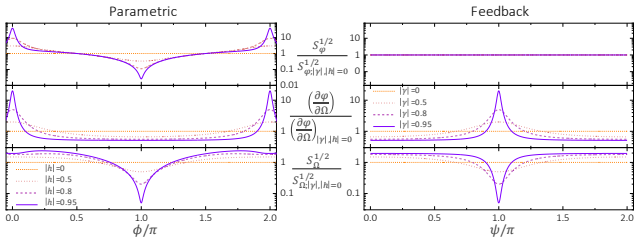

**Figure 3 Thermomechanical noise limited |** Effect of both parametric pumping (left) and feedback (right) on $S_\varphi^{1/2}$, $\left(\frac{\partial \varphi}{\partial \Omega}\right)_{\Omega=0}$, and $S_\Omega^{1/2}$, relative to the case with no pumping and no feedback. In this case, the noise source is purely thermomechanical and the driving amplitude, $|g|$, is held constant. $S_\varphi$ does not depend on $\psi$ whereas it depends on $\phi$, showing a minimum for $\phi = \pi$. The slope of the phase shows a maximum for $\psi = \pi$ and $\phi = 0$, whereas a minimum for $\phi = \pi$ and $\psi = 0$. $S_\Omega$ shows a minimum for both cases around $\psi = \pi$ and $\phi = \pi$. This improvement would keep increasing with both parametric ($|h|$) and/or feedback ($|\gamma|$) strength until reaching the threshold. However, a physical system will exhibit nonlinearities before the threshold, which are not accounted for in this figure.

Even if the results shown in Figure 3 are interesting, they are not representative of physical systems since this analytical study involved solely linear terms for clarity. However, it is known that any system shows nonlinearities after a certain threshold. In order to account for this without performing a full nonlinear analysis of the system, we can replot the results of Figure 3 while holding the amplitude of vibration constant and equal to the critical amplitude (onset of nonlinearity)[23]. The results of this approach can be seen in Figure 4, illustrating a behavior that is radically different from the previously shown results in Figure 3.

If the amplitude is kept constant, allowing the intensity of the drive, $|g|$ to change for every case, we can clearly see how the noise in the resonator's phase ($S_\varphi$) gets worse around the amplification region ($\psi = \pi$), whereas it reaches a minimum when the feedback attenuates ($\psi = 0$). Of course, this means that the value of the drive in either case will be very different, with the drive in the latter case being much larger than in the former ($|g|_{\psi=0} \gg |g|_{\psi=\pi}$). Combining this with the changes in the phase slope, it is possible to see that they compensate in the two extreme cases, and the frequency noise is the same as the one without any feedback. In the remaining cases, the frequency noise is larger than the one obtained without any feedback.

Analyzing the parametrically pumped case, we observe that the noise in the resonator's phase reaches a minimum within the amplification stage ($\phi = \pi$) and a maximum within the attenuation stage ($\phi = 0$). This is due to the squeezing, i.e., the non-symmetrical distribution around the average of the noise. Combining this with the phase slope, we reach a similar conclusion as for the direct feedback case: the frequency noise is the same as in the case without pump for $\phi = 0, \pi$ but it is larger otherwise. The former point is a result that has been reached repeatedly in the literature. However, some controversy exists since some works have been claiming that an improvement is possible by using these techniques. Here, we actually show that for most values of the phase delay, not only does the situation not improve, but it actually worsens.

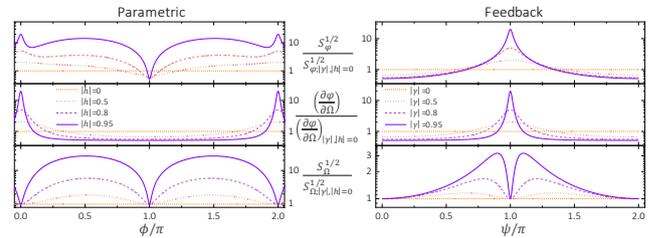

**Figure 4 Thermomechanical noise limited. Nonlinear cap. |** Same content as in Figure 3, except the amplitude is set to be equal to the critical value, i.e., $|g|$ is adjusted automatically for each phase case. In this case, $S_\varphi$ increases close to $\psi = \pi$ (amplifying phase), and close to $\phi = 0$ (attenuation phase), whereas it exhibits a minimum at $\phi = \pi$. The slope of the phase follows a similar trend as the one described in Figure 3. Finally, when looking at $S_\Omega$, it is impossible to find a phase, where there is an improvement compared to the case without any feedback or pumping. $S_\Omega$ can only get worse with increasing feedback or parametric pumping.

### 3.1 Amplifier noise

The situation is slightly different when we consider that it is not thermomechanical noise, but amplifier noise that limits many resonant sensors. Amplifier noise is a generic term that includes transduction noise or, even more generally, any noise in the measurement that is not thermomechanical. Amplifier noise does not directly affect the dynamics of the system but it simply adds to the output noise variance in an isotropic manner. Therefore it is straightforward to calculate its effect onto the parametric pump case. On the other hand, its effect on the direct feedback is more involves: it will actually depend on whether the feedback is generated with the amplifier noise (most common case) or not. In the former case, for example, the feedback will also include a noisy force in the system, which magnitude will depend on the feedback gain and phase. In Figure 5, we plot the estimation of frequency noise for all 3 cases (parametric pumping, noiseless feedback, and noisy feedback) assuming thermomechanical noise negligible, for constant drive ($|g|$), and constant amplitude ($\tilde{x}_c$).

We can observe that for direct feedback the optimum operational point is always located at $\psi = \pi$, which is the amplification stage. At this point, the slope is maximized and, for the case of constant drive, the amplitude is much larger which implies a smaller noise in the resonator's phase. Interestingly, when considering the more likely situation of a noisy feedback (i.e., a feedback that is generated using a signal that has some transducer or amplifier noise), and keeping the amplitude constant, $S_\Omega$ reduces around the amplification region for small values of $|\gamma|$. This improvement, however, saturates when reaching $|\gamma| = 0.5$ only to later go back to the value of $S_\Omega$ without feedback (when $|\gamma| \to 1$). It is important to remember that the *actual* minimum of this frequency noise will depend on the relative intensity of amplifier noise and thermomechanical noise, since the latter is amplified by the direct feedback at $\psi = \pi$, and at some point it will equate and surpass the amplifier noise.

In Figure 5 we also show that, in the case of parametric pumping, the result depends on whether we keep the drive or vibration amplitude constant. In the former case, we obtain a minimum in the frequency noise for $\phi = 0, \pi$, due to both the changes in the slope and the amplitude. However, in the case when the vibration amplitude is held constant, and contrary to what one would expect, the frequency noise exhibits the largest improvement when the attenuation phase is used, since the slope of the phase is much larger. In this case, however, the driving amplitude needs to be larger in order to compensante for the attenuation caused by the parametric pumping. As for the case of direct feedback, these results are only valid when the amplifier noise dominates. In reality, the results depend on the relative intensity of the thermomechanical and amplifier noise.

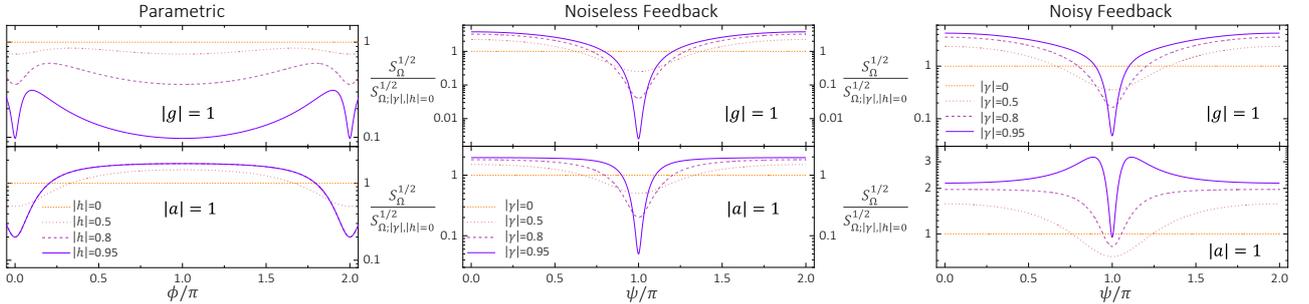

**Figure 5 Amplifier noise limited |** $S_\Omega$ of a resonator with parametric pumping (left), noiseless linear feedback (center) and noisy linear feedback (right) relative to the case with no pump nor feedback. On the top row we show the case where the drive ($|g|$) is fixed (to be compared to (7)) and on the bottom row we show the case where the amplitude of motion ($|a|$) is fixed (to be compared to (7)). In all cases here, the limiting noise is not thermomechanical, but amplifier noise. The results for the linear feedback case, as expected, show an improvement when using an amplifying feedback phase $\psi = \pi$. Importantly, in the more likely case of a noisy feedback and operating at constant amplitude, the feedback slightly improves performance for small $|\gamma|$ but then it goes back to the performance without feedback. In the case of parametric pump, when the driving force is kept constant, an improvement of the frequency noise can be observed no matter the phase, being maximum for $\phi = 0, \pi$. When the amplitude of motion is kept constant, $S_\Omega$ improves for the attenuation cases, $\phi = 0$. This somewhat counterintuitive result is explained by the dependence of the phase slope on the parametric pump, as seen in Figure 1.

## 4. Conclusion

In this paper, the effect of parametric pumping and direct feedback on a mechanical resonator is investigated in the presence of thermomechanical noise, amplifier noise, and direct forcing. We first analyze the noiseless solution, observing in the case of parametric pumping that the phase slope decreases for the amplification stage, as opposed to the direct feedback case, where both the amplitude and phase slope are maximized concurrently. This difference is also visible in the solution with noise, where the parametrically pumped case shows noise distributions that are anisotropic around the average values. This is, to our knowledge, the first time where this solution is shown, and we confirm it by numerical integration and via experimental measurements. We then analyze the effect of this noise distribution on the frequency noise and observe that, when the driving force is constant, the frequency noise can be minimized by using either direct feedback or parametric pumping. However, when operating at the critical amplitude, the frequency noise cannot be improved compared to the case without feedback nor pumping. The situation is different for a system that is limited by amplifier noise instead of thermomechanical noise. In that case, the frequency noise can be reduced at several phases. The most interesting result is obtained when operating at the critical amplitude with parametric pumping, where we find that the frequency noise is minimized in the attenuation case, contrary to established concepts. We believe that the results within this paper have important implications for the field of nano- and micromechanical sensing.

## Acknowledgements

L.G.V. and Z.M. thank financial support to the Swiss SNF (grant PP00P2_170590). J.M.L.M. was supported by the NSF (grant CMMI-1662464), with partial support from the NDSEG Fellowship and the E.K. Potter Stanford Graduate Fellowship. O.Z. acknolwledges financial support from the Swiss SNF (grant PP00P2_163818). T.L.H. acknowledges financial support from the Swiss SNF (grant CRSII5_177198). Fabrication was performed in the nano@Stanford labs, supported by the NSF (grant ECCS-1542152) and DARPA PRIGM Program.

## References

[1] Naik, A. K.; Hanay, M. S.; Hiebert, W. K.; Feng, X. L.; Roukes, M. L., Towards single-molecule nanomechanical mass spectrometry. *Nature Nanotechnology* **2009,** *4* (7), 445-450.
[2] Whiting, J. J.; Lu, C. J.; Zellers, E. T.; Sacks, R. D., A portable, high-speed, vacuum-outlet GC vapor analyzer employing air as carrier gas and surface acoustic wave detection. *Analytical Chemistry* **2001,** *73* (19), 4668-4675.
[3] Menon, A.; Zhou, R. N.; Josse, F., Coated-quartz crystal resonator (QCR) sensors for on-line detection of organic contaminants in water. *Ieee Transactions on Ultrasonics Ferroelectrics and Frequency Control* **1998,** *45* (5), 1416-1426.
[4] Mamin, H. J.; Poggio, M.; Degen, C. L.; Rugar, D., Nuclear magnetic resonance imaging with 90-nm resolution. *Nature Nanotechnology* **2007,** *2* (5), 301-306.
[5] Ramos, D.; Tamayo, J.; Mertens, J.; Calleja, M.; Villanueva, L. G.; Zaballos, A., Detection of bacteria based on the thermomechanical noise of a nanomechanical resonator: origin of the response and detection limits. *Nanotechnology* **2008,** *19* (3).
[6] Cleland, A. N.; Roukes, M. L., A nanometre-scale mechanical electrometer. *Nature* **1998,** *392* (6672), 160-162.
[7] Larsen, T.; Schmid, S.; Gronberg, L.; Niskanen, A. O.; Hassel, J.; Dohn, S.; Boisen, A., Ultrasensitive string-based temperature sensors. *Applied Physics Letters* **2011,** *98* (12).
[8] Larsen, T.; Schmid, S.; Villanueva, L. G.; Boisen, A., Photothermal Analysis of Individual Nanoparticulate Samples Using Micromechanical Resonators. *Acs Nano* **2013,** *7* (7), 6188-6193.
[9] Zhang, X. C.; Myers, E. B.; Sader, J. E.; Roukes, M. L., Nanomechanical Torsional Resonators for Frequency-Shift Infrared Thermal Sensing. *Nano Lett* **2013,** *13* (4), 1528-1534.
[10] Ekinci, K. L.; Roukes, M. L., Nanoelectromechanical systems. *Review of Scientific Instruments* **2005,** *76* (6).
[11] Cassella, C.; Segovia-Fernandez, J.; Piazza, G.; Cremonesi, M.; Frangi, A., Reduction of Anchor Losses by Etched Slots in Aluminum Nitride Contour Mode Resonators. *2013 Joint European Frequency and Time Forum & International Frequency Control Symposium (Eftf/Ifc)* **2013,** 926-929.
[12] Segovia-Fernandez, J.; Cremonesi, M.; Cassella, C.; Frangi, A.; Piazza, G., Anchor Losses in AlN Contour Mode Resonators. *J Microelectromech S* **2015,** *24* (2), 265-275.
[13] Richter, A. M.; Sengupta, D.; Hines, M. A., Effect of surface chemistry on mechanical energy dissipation: Silicon oxidation does not inherently decrease the quality factor. *Journal of Physical Chemistry C* **2008,** *112* (5), 1473-1478.
[14] Villanueva, L. G.; Schmid, S., Evidence of Surface Loss as Ubiquitous Limiting Damping Mechanism in SiN Micro- and Nanomechanical Resonators. *Phys Rev Lett* **2014,** *113* (22).
[15] Rugar, D.; Grutter, P., Mechanical Parametric Amplification and Thermomechanical Noise Squeezing. *Phys Rev Lett* **1991,** *67* (6), 699-702.
[16] Prakash, G.; Raman, A.; Rhoads, J.; Reifenberger, R. G., Parametric noise squeezing and parametric resonance of microcantilevers in air and liquid environments. *Review of Scientific Instruments* **2012,** *83* (6), 065109.
[17] Wilson, D. J.; Sudhir, V.; Piro, N.; Schilling, R.; Ghadimi, A.; Kippenberg, T. J., Measurement-based control of a mechanical oscillator at its thermal decoherence rate. *Nature* **2015,** *524* (7565), 325-329.
[18] Sansa, M.; Sage, E.; Bullard, E. C.; Gely, M.; Alava, T.; Colinet, E.; Naik, A. K.; Villanueva, L. G.; Duraffourg, L.; Roukes, M. L., Frequency fluctuations in silicon nanoresonators. *Nature Nanotechnology* **2016,** *in press*.
[19] Tamayo, J.; Humphris, A. D. L.; Malloy, A. M.; Miles, M. J., Chemical sensors and biosensors in liquid environment based on microcantilevers with amplified quality factor. *Ultramicroscopy* **2001,** *86* (1–2), 167-173.
[20] Cleland, A. N., Thermomechanical noise limits on parametric sensing with nanomechanical resonators. *New Journal of Physics* **2005,** *7* (1), 16.
[21] Buks, E.; Yurke, B., Mass detection with a nonlinear nanomechanical resonator. *Phys Rev E* **2006,** *74* (4 Pt 2), 046619.
[22] Villanueva, L. G.; Kenig, E.; Karabalin, R. B.; Matheny, M. H.; Lifshitz, R.; Cross, M. C.; Roukes, M. L., Surpassing Fundamental Limits of Oscillators Using Nonlinear Resonators. *Phys Rev Lett* **2013,** *110* (17), 177208.
[23] Lifshitz, R.; Cross, M. C., Nonlinear Dynamics of Nanomechanical and Micromechanical Resonators. In *Reviews of Nonlinear Dynamics and Complexity*, Wiley-VCH Verlag GmbH & Co. KGaA: 2009; pp 1-52.
[24] Kenig, E.; Cross, M. C., Eliminating 1/f noise in oscillators. *Phys Rev E* **2014,** *89* (4).
[25] Kenig, E.; Cross, M. C.; Villanueva, L. G.; Karabalin, R. B.; Matheny, M. H.; Lifshitz, R.; Roukes, M. L., Optimal operating points of oscillators using nonlinear resonators. *Phys Rev E* **2012,** *86* (5).
[26] Vinante, A.; Falferi, P., Feedback-Enhanced Parametric Squeezing of Mechanical Motion. *Phys Rev Lett* **2013,** *111* (20).

[27] Miller, J. M. L.; Bousse, N. E.; Heinz, D. B.; Kim, H. J. K.; Kwon, H. K.; Vukasin, G. D.; Kenny, T. W., Thermomechanical-Noise-Limited Capacitive Transduction of Encapsulated MEM Resonators. *J Microelectromech S* **2019,** *28* (6), 965-976.

[28] Miller, J. M. L.; Bousse, N. E.; Shin, D. D.; Kwon, H.; Kenny, T. W. In *Signal Enhancement in MEM Resonant Sensors Using Parametric Suppression*, 2019 20th International Conference on Solid-State Sensors, Actuators and Microsystems & Eurosensors XXXIII (TRANSDUCERS & EUROSENSORS XXXIII), 23-27 June 2019; 2019; pp 881-884.

[29] Miller, J. M. L.; Ansari, A.; Heinz, D. B.; Chen, Y. H.; Flader, I. B.; Shin, D. D.; Villanueva, L. G.; Kenny, T. W., Effective quality factor tuning mechanisms in micromechanical resonators. *Appl Phys Rev* **2018,** *5* (4).